# Application of Machine Learning Techniques in Aquaculture

Akhlaqur Rahman[1] and Sumaira Tasnim[2]

[1]Department of Electrical and Electronic Engineering Uttara University, Bangladesh
[2]School of Engineering, Deakin University, Australia

*ABSTRACT:*
In this paper we present applications of different machine learning algorithms in aquaculture. Machine learning algorithms learn models from historical data. In aquaculture historical data are obtained from farm practices, yields, and environmental data sources. Associations between these different variables can be obtained by applying machine learning algorithms to historical data. In this paper we present applications of different machine learning algorithms in aquaculture applications.

## 1. INTRODUCTION

Aquaculture refers to the farming of aquatic organisms such as fish and aquatic plants. It involves cultivating freshwater and saltwater populations under controlled conditions. Use of sensor technologies to monitor the environment where aquaculture operations take place is a recent trend. The sensors collect data about the aquaculture environment that are using by farm managers for decision making purposes.

The literature states a number of activities related to decision support systems in Aquaculture farm operations. A number of decision support systems have been developed for. Some of them use machine learning and methods and other do not. For the sake of completeness we briefly discuss these other methods first and detail the machine learning based methods in the following section.

Bourke et al. [1] developed a framework where real-time water quality indicators, as well as operational information were displayed and their impact on survival rate, biomass and production failure of aquaculture species were evaluated. Wang et al. [2] developed an early warning system for dangerous growing conditions. Padala and Zilber [3] used expert system generated rules to reduce stock loss and increase size and quality of yields. Ernst et al. [4] focused on managing hatchery production using rules and calculations of physical, chemical and biological processes. Silvert [5] developed a scientific model to evaluate the environmental impact. Halide et al. [6] used rules that are hand-crafted from domain experts.

## 2. MACHINE LEARNING METHODS IN AQUACULTURE APPLICATIONS

In this section we discuss the different applications where machine learning methods are applied in the aquaculture domain.

### 2.1 Shellfish Farm Closure Prediction

Consumption of contaminated shellfish can cause severe illness and even death in humans. The authors in [7]–[9] developed a sensor network based approach to predict the contamination event using machine learning methods. In [7] the authors presented a machine learning methods to obtain a balance between farm closure and farm opening events. The authors in [8] presented a feature ranking algorithm to identify the most influential cause of closure. In [9] the authors adopted time series machine learning approaches like PCA (Principal Component Analysis) and ACF (Auto Correlation Function) to predict the closure event.

### 2.2 Algae Bloom Prediction

Algae organisms grow widely throughout in the world. They provide food and shelter to other organisms. When the growth of algae is excessive, it can cause oxygen depletion in water and kill fish. The authors in [10]–[12] developed machine learning based methods to predict algae growth/bloom. The authors in [10] extracted a set of rules from data gathered by sensor networks to find associations between environmental variables and algae growth. The authors in [11] designed an ensemble method to find the relevant environmental variables responsible for algae growth and predict the growth. The authors in [12] developed machine learning methods to predict the propagation of algae patches along the waterway.

### 2.3 Missing Values Estimation

Sensor networks deployed in the field to monitor aquaculture environments often suffer failure due to sensor bio-fouling, communication failure etc. This results in missing sensor readings that is required by the machine learning based decision making systems. This results in a requirement to deal with missing sensor values. The authors in [13] and [14] designed a multiple classifier based method to deal with missing sensor data. Instead of imputing missing data, prediction of events is based on available sensor readings. Ensemble approach is shown to perform better than imputation methods. The authors assume equal weights for all features which may not be true in real world scenarios [15][16].

### 2.4 Model Relocation

In order to train machine learning methods it is required to provide historical labeled data. However when a farmer aims to setup a new aquaculture firm, historical farming data is not available for that location. The authors in [17] provided a guideline to utilize machine learning models trained for a different location to a new location based on similarity between the two locations. Results show that model relocation can significantly reduce the shortcoming generated from data unavailability for a particular location.

### 2.5 Benthic Habitat Mapping

In order to get an understanding of the aquaculture habitats on seafloor the researchers sometimes send Autonomous Underwater Vehicles to collect seafloor images. The images are later visually analyzed by the researchers to produce the habitat map in a region. The authors in [18] developed an image processing and machine learning based method to automatically produce habitat maps from seafloor images. The results presented in the paper showed significant accuracy in obtaining correct habitat maps.

### 2.6 Sensor Data Quality Assessment

Data produced by the sensors are sometimes faulty. The decisions based on faulty sensor reading will result in wrong conclusion. In [19][20] the authors have presented a novel ensemble classifier approach for assessing the quality of sensor data. The base classifiers are constructed by random under–sampling of the training data where the sampling process is guided by clustering. The inclusion of cluster based under–sampling and multi–classifier learning has shown to improve the accuracy of quality assessment.

### 3. CONCLUSIONS

In this paper we have presented a review of machine learning methods in the aquaculture space. In future we aim to extend some of these methods n time series machine learning methods.